\documentclass [11pt,a4,twoside]{book}
\usepackage[dvips]{graphicx}
\usepackage{dropping}
\usepackage{rotating}
\usepackage{lscape}
\usepackage{longtable}
\usepackage{deluxetable}
\def\hel{{\tt HELLAS2XMM}}
\def\odf{{\tt HELLAS2XMM 1dF}}

\def\chandra{{\it Chandra}}
\def\xmm{XMM--{\it Newton}}

\def\pn{\par\noindent} % Non indenta un nuovo paragrafo
\def\simlt{\mathrel{\spose{\lower 3pt\hbox{$\mathchar"218$}}
     \raise 2.0pt\hbox{$\mathchar"13C$}}}
\def\simgt{\mathrel{\spose{\lower 3pt\hbox{$\mathchar"218$}}
     \raise 2.0pt\hbox{$\mathchar"13E$}}}
\def\ls{\mathrel{\hbox{\rlap{\hbox{\lower4pt\hbox{$\sim$}}}\hbox{$<$}}}}
\def\gs{\mathrel{\hbox{\rlap{\hbox{\lower4pt\hbox{$\sim$}}}\hbox{$>$}}}}
\def\lesssim{\mathrel{\hbox{\rlap{\hbox{\lower4pt\hbox{$\sim$}}}\hbox{$<$}}}}
\def\gtrsim{\mathrel{\hbox{\rlap{\hbox{\lower4pt\hbox{$\sim$}}}\hbox{$>$}}}}
\def\gsimeq{\mathrel{\hbox{\rlap{\hbox{\lower4pt\hbox{$\sim$}}}\hbox{$>$}}}}
\def\lsimeq{\mathrel{\hbox{\rlap{\hbox{\lower4pt\hbox{$\sim$}}}\hbox{$<$}}}}
\def\spose#1{\hbox to 0pt{#1\hss}}

\def\erg{${\rm erg ~cm^{-2} ~s^{-1}}$}
\def\cgs{${\rm erg ~cm^{-2} ~s^{-1}}$}
\def\due	{$^{-2}$}

\newcommand{\msigma}{$M_{\bullet}$--$\sigma$\ }

\renewcommand{\thefigure}{\arabic{chapter}.\arabic{figure}}
\renewcommand{\thetable}{\arabic{chapter}.\arabic{table}}
\renewcommand{\thepage}{\arabic{page}}
\def\fnum@figure{{\sf Figure \thefigure}}
\def\fnum@table{{\sf Table \thetable}}
\def\@chapapp{Chapter}
\def\s0{\setcounter{equation}{0}\setcounter{table}{0}\setcounter{figure}{0}}
\def\thebibliography#1{\section*{Bibliography} 
 \list
 {{\sf [\arabic{enumi}]}}{
\settowidth\labelwidth{[#1]}\leftmargin\labelwidth
 \advance\leftmargin\labelsep
 \usecounter{enumi}}
 \def\newblock{\hskip .11em plus .33em minus .07em}
 \sloppy\clubpenalty4000\widowpenalty4000
 \sfcode`\.=1000\relax}

\begin{document}
\baselineskip 6.5mm
\vbadness=10000\hbadness=10000\tolerance=10000
%%%%%%%%%%%%
% Chapters
%%%%%%%%%%%%
% here the title 
\renewcommand{\thepage}{\Roman{page}}
\pagestyle{empty}
\font\grandebf=cmbx10 scaled\magstep3
\font\bigbf=cmbx10 scaled\magstep2
\font\pippo=cmbx10 scaled\magstep1
\parindent 0pt
\centerline {\grandebf Universit\`a degli Studi di Bologna}
\vskip 0.2 truecm
\hrule
\vskip 0.1 truecm
\hrule
\vskip 0.5 truecm
\centerline {FACOLT\`A DI SCIENZE MATEMATICHE, FISICHE E NATURALI}
\vskip 0.3 truecm
\centerline {Dipartimento di Astronomia}
\vfill
\vskip 1.4 truecm 
\centerline {\grandebf PHYSICS AND EVOLUTION OF} 
\vskip 0.8 truecm
\centerline {\grandebf OBSCURED X--RAY SOURCES:}
\vskip 0.8 truecm
\centerline {\grandebf A MULTIWAVELENGTH APPROACH}
\vfill
\vskip 1.2 truecm
\vskip 0.3 truecm
\centerline {Tesi di dottorato di}
\vskip 0.3 truecm
\centerline {\pippo MARCELLA BRUSA}
\vskip 2.5 truecm
{Tutore:}
\vskip 0.2 truecm
{\pippo Prof. Bruno Marano}
\vskip 1. truecm
{Co-Relatore:}
\vskip 0.2 truecm
{\pippo Dr. Andrea Comastri}
\vskip 1. truecm
{Coordinatore:}
\vskip 0.2 truecm
{\pippo Prof. Gabriele Giovannini}

\vskip 1.4 truecm
\vskip 0.3 truecm
\hrule
\vskip 0.1 truecm
\hrule
\vskip 0.2 truecm
\centerline {DOTTORATO DI RICERCA IN ASTRONOMIA}
\vskip 0.2 truecm
\centerline{XVI CICLO [2000--2003]}

\cleardoublepage

\setcounter{page}{1}
% here the index
\tableofcontents
\cleardoublepage
%%%%%%%%%%%%%%%%%%%%%%%%%%%%%%%%%%%%%%%%%%
\renewcommand{\thepage}{\arabic{page}}
\setcounter{page}{1}
\large

\clearpage
\markboth{\sc \ }{\sc Introduction}

\addcontentsline{toc}{chapter}{\bf Introduction}
%\addtocontents{toc}{\bf Introduction}
\vspace*{1.6truecm}
\begin{flushleft}
{\Huge \bf Introduction}
\end{flushleft}
\vspace {1.0truecm}
\setcounter{chapter}{1}
\setcounter{section}{0}
\setcounter{figure}{0}
\setcounter{table}{0}
\setcounter{equation}{0}

%%%%%%%%%%%%%%%%%%%%%%%%%%%%%%%%%%%%%%%%%%%%%%%%%%%%%%%%%%%%%%%
\par\noindent
{\sf \dropping[8pt]{3}{O}ne of the primary goals of observational 
cosmology is to trace star formation and nuclear activity along 
with the mass assembly history of galaxies as a function of redshift 
and environment.
Theoretical arguments suggest that there is a fundamental link between the
assembly of Black Holes and the formation of spheroids in galaxy
halos. 
The tight relation observed in local galaxies between the black
holes mass and the velocity dispersion (the M$_{\rm BH}$-$\sigma$ 
relation) %that is a measure of the gravitational potential well 
and the fact that the locally inferred black hole mass density
appears to be broadly consistent with the mass accreted during the
quasar phase further support the idea that the nuclear activity,
the growth of the black holes and spheroid formation are all closely
linked.\\    
Thus, it is clear that a detailed investigation of the formation and
evolution of Active Galactic Nuclei (AGN) over a wide range
of redshifts, and the comparison with galaxy evolution, 
could provide information about the link between nuclear activity, 
star formation and the galaxy assembly that seem to co-exist in the 
early Universe. In this framework, the hard X--ray band is by far the 
cleanest one where to study the history of accretion in the Universe, being
the only band in which accretion processes dominate the 
cosmic background. 

%%%%%%%%%%%%%%%%%%%%%%%%%%%%%%%%%%%%%%%%%%%%%%%%%%%%%%%%%
\section*{The X--ray background}
\addcontentsline{toc}{section}{The X--ray background}

The cosmic X--ray background (XRB) was discovered at the dawn of the
X--ray astronomy: 
during the first successful rocket flight launched to study
the X--ray emission from the Moon, the presence of a residual 
diffuse emission was also ``serendipitously'' revealed 
(Giacconi et al. 1962).
The lack of any correlation with the galactic latitude, and a 
dipole anisotropy consistent with that of the dipole component of the 
Cosmic Microwave Background (CMB; Shafer \& Fabian 1983) 
strongly argued from the beginning in 
favour of a cosmological origin of this extragalactic 
background radiation. \\ 
\pn
The first broad--band measurements of the XRB spectrum was obtained 
at the end of the 70's with the HEAO--1 satellite. 
Marshall et al. (1980) showed that the HEAO--1 data were very
well fitted by a 40 keV thin thermal bremsstrahlung model,
approximated below 15 keV with a simple power--law spectral 
function ($F(E)\propto E^{-\Gamma}$) with photon index $\Gamma\sim1.4$. 
It was therefore quite natural to hypothesize the presence of a truly
diffuse hot Inter Galactic Medium (IGM) with a characteristic 
temperature of kT=40(1+z) keV, as originally proposed by 
Field and Perrenod (1977).  \\
However, a reasonable extrapolation of the X--ray properties and optical counts
of extragalactic sources led to the conclusion that discrete sources 
could contribute significantly to the XRB (Schmidt \& Green 1986).  
Moreover, Giacconi \& Zamorani (1987) have subsequently shown that, 
once the contribution estimated from known AGN is removed, 
the residual background spectrum is too flat to be interpreted 
in terms of an optically thin bremsstrahlung from a diffuse IGM.\\
This hypothesis has been then entirely discarded with the 
results from the COBE satellite (Mather et al. 1990).  
A diffuse, hot (T $\gtrsim$ 10$^{8}$~K) IGM should give rise to evident 
high-frequency distortions in the CMB spectrum through inverse-Compton 
scattering, which have not been observed by COBE; 
this implies that the contribution of hot  gas to the XRB is lower 
than 0.01\% (Wright et al. 1994).   \\
%%%%%%%%%%%%%%%%%%%%%%%%%%%%%%%
\begin{figure}[!t]
\includegraphics[width=\textwidth]{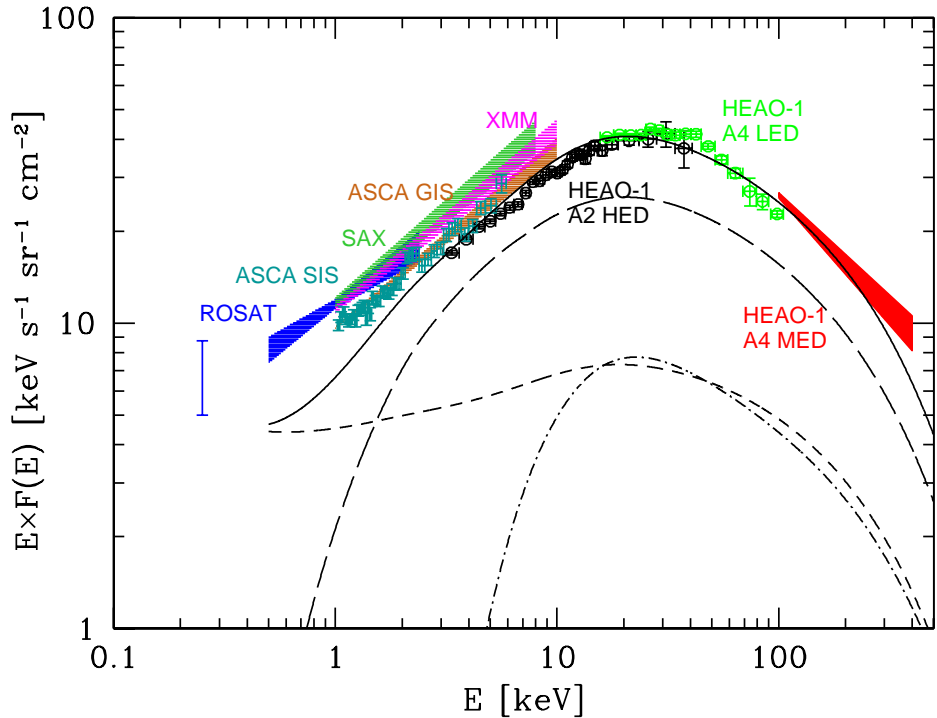}
\caption{The AGN contribution (solid line) to the
XRB spectral energy density as measured by different instruments (labeled).
The contribution of unobscured (N$_H<10^{21}$ cm$^{-2}$, dashed line),
Compton thin (N$_H=10^{21}-10^{24}$ cm$^{-2}$, long--dashed line) 
and Compton thick (N$_H)>10^{24}$ cm$^{-2}$, dot--dashed line)
AGN is also shown. [From Comastri (2004)].} 
\label{xrb_fit}
\end{figure}                                                                
%%%%%%%%%%%%%%%%%%%%%%%%%%%%%%%%%%%
\pn
As a consequence, the only viable alternative for the XRB origin 
remained the superposition of discrete sources, and the most likely 
candidates appeared immediately to be AGN. \\
The most serious problem with the discrete-source origin for the XRB
remained the so-called ``spectral paradox'' (Boldt 1987): 
at that time there were already stringent evidences that 
the AGN X--ray continuum in the 2--10 keV band is well described 
by a power--law with index $\Gamma$ $\simeq$ 1.7--1.9 (Mushotzky et al. 
1984; Turner \& Pounds 1989), too soft to fit the value observed for 
the hard XRB in the same energy band ($\Gamma$ $\sim$ 1.4). \\
Setti \& Woltjer (1989) showed that, on the basis of Unification 
Schemes of AGN, strong X--ray absorption 
is naturally predicted for the sources optically classified as narrow 
line, Type 2 objects\footnote{In the zero--th order Unification Schemes, 
there is a correspondence between X--ray and optical absorption: Type 2 
narrow--lined objects are X--ray obscured, and Type 1 broad--lined objects 
are not. The underlying continuum is intrinsically the same, and the 
differences in the optical and X--ray spectra are only due to orientation 
effects (e.g. Antonucci 1993).}, 
and that the amount of obscuring matter along the line of
sight, measured in equivalent neutral hydrogen column (N$_H$),
can be even in excess than 10$^{24}$ cm$^{-2}$. 
As a consequence, the resulting X--ray spectra will peak at high energies 
(from a few keV up to $\gs10$ keV, depending from the N$_H$) and the net 
result is a flatter power--law spectral index. 
The ``spectral paradox'' could be therefore theoretically 
solved by assuming that the XRB is due to the superposition of absorbed and 
unabsorbed objects, with the same intrinsic steep ($\Gamma\sim1.8$) power--law 
continuum. \\
Following these indications, several authors have refined and 
developed population--synthesis models able to reproduce the 
XRB spectral shape and intensity (e.g. Madau, Ghisellini \& Fabian 
1994; Comastri et al. 1995,2001; Gilli, Salvati \& Hasinger 2001), assuming a wide range
of redshifts (i.e. evolution) and column densities\footnote{If 
the X--ray obscuring matter has a column density which is equal or larger 
than the inverse of the Thomson cross--section 
($N_H \ge \sigma_T^{-1} \simeq 1.5 \times 10^{24}$ cm$^{-2}$) the source is called, 
by definition, ``Compton Thick''.} of the absorbing matter
in the range N$_H=10^{21}-10^{25}$ cm$^{-2}$. 
In Fig.~\ref{xrb_fit} a compilation of XRB measurements in different
energy bands is shown along with the best--fit model from Comastri
et al. (2001). The contribution of unobscured (dashed line),
mildly obscured (N$_H=10^{21}-10^{24}$ cm$^{-2}$, Compton Thin) 
and heavily obscured (N$_H>10^{24}$ cm$^{-2}$, Compton Thick) sources is also reported.\\
The combined fit of other observational constraints in addition to 
the XRB spectral shape (in primis the number counts, and the redshift
and absorption distributions in different energy ranges) is needed 
to build a self--consistent model and discriminate among the 
assumptions adopted for the cosmological evolution of the discrete sources.
In particular, the key parameter turned out to be the evolution of obscured 
(Type 2) sources.  
As an example, the baseline AGN synthesis model adopt a simple
Pure Luminosity Evolution (PLE) and is based on the zero--th order
AGN Unification Schemes (Antonucci 1993); 
this implies that 1) the evolution of obscured 
AGN is {\it assumed} to follow that of unobscured AGN, and 2) the
existence of a population of high--luminosity, highly obscured 
quasars (the so--called QSO2) is {\it naturally} postulated 
(Comastri et al. 1995).
Indeed, it has been subsequently shown that QSO2 are {\it necessary} 
in reproducing the 2--10 keV source counts at relatively bright 
fluxes ($\simeq 10^{-13}$ \cgs; Gilli et al. 2001; Comastri et al. 2001); 
however, despite intensive optical searches, 
these narrow--line high--redshift objects appear to be elusive, suggesting a 
space density and evolution different from that expected from unified
schemes and calling for substantial revisions of the main assumptions 
of XRB baseline models. 

%%%%%%%%%%%%%%%%%%%%%%%%%%%%%%%%%%%%%%%%%%%%%%%%%%%%%%%%%%%%%%%%%%%
\section*{Hard X--ray surveys and ``new'' observational constraints}
\addcontentsline{toc}{section}{Hard X--ray surveys and new observational constraints} 
A large number of multiwavelength projects have
been specifically designed in the last years 
%to better understand the physical nature of
to investigate the evolutionary properties of AGN. \\
The results from hard X--ray surveys carried out at the end of the 
90's with ASCA (e.g. the ASCA {\tt GIS} survey, Cagnoni et al. 1998;
the ASCA {\tt MSS} and {\tt LSS} Surveys, Ueda et al. 1999; Akiyama 
et al. 2002) and BeppoSAX (the {\it Beppo}SAX {\tt HELLAS} survey, 
Fiore et al. 2001; Vignali 2001) constituted the first piece of evidence 
that AGN synthesis models do indeed work. However, the resolved
fraction of the XRB was still too low ($\sim 25-30$\%) to
quantitatively constrain all the models parameters; in particular,
sampling the bright X--ray fluxes ($>10^{-13}$ \cgs) previous X--ray surveys were
strongly biased towards luminous, high--redshift sources. \\ 
The superb capabilities of the detectors on-board the \chandra\ and
\xmm\ X--ray satellites have opened up new frontiers in testing
these predictions down to limiting fluxes where 
the entire spectral energy density in the 2--10 keV band is
expected to be resolved. 
Thanks to the deep X--ray surveys carried out 
in the \chandra\ Deep Field North (CDFN, Brandt et al. 2001a,
Alexander et al. 2003), \chandra\ Deep Field South 
(CDFS, Giacconi et al. 2001) and the Lockman Hole (Hasinger et al. 2001) 
the X--ray sky is now probed down to a 2--10 keV flux limit of about 
$2\times 10^{-16}$ erg cm$^{-2}$ s$^{-1}$ and a fraction as
large as 80-90\% of the diffuse XRB is resolved into
discrete sources (Mushotzky et al.~2000; Hasinger et al. 2001; 
Rosati et al. 2002; Alexander et al. 2003), the bigger uncertainty
being the XRB normalization (Barcons, Mateos, \& Ceballos 2000).\\
Moreover, hard X--ray surveys have proven to be very efficient 
to uncover obscured accreting black holes, confirming, at least 
qualitatively, the predictions of standard models, 
in which the 2-10 keV XRB is mostly made by the
superposition of obscured and unobscured AGN. 
% (e.g. Setti \& Woltjer 1989, Comastri et al. 1995, 2001). \\
In particular, the recent findings are in very good agreement with 
the main predictions of XRB synthesis models for the source counts 
in different X--ray bands (see Moretti et al. 2003 for a comprehensive 
compilation of recent data); moreover, the observed average spectrum of the sources 
detected down to $10^{-16}$ \cgs\ now exactly matches that of the
XRB ($<\Gamma>\sim1.4$, Tozzi et al. 2001a,b).\\ 
Deep, pencil beam surveys, albeit extremely important, 
explore only a limited region of the luminosity--redshift plane, 
being strongly biased against the brightest (and rare) objects. 
Sizable samples of objects detected at the bright X--ray fluxes
($\gtrsim 10^{-14}$ \cgs) over an area of the order of a few  square degrees
are needed to homogeneously cover the Hubble diagram and
to obtain a well--constrained luminosity function with a similar number 
of sources per luminosity decade and per redshift bin.
The detailed study of the nature of the hard X--ray source population,
is indeed pursued complementing deep pencil beam observations 
with shallower, larger area surveys.  
In the last few years several projects with both \chandra\ and
\xmm\ have already started, with the aim of surveying from few to
several tens degrees of the hard X--ray sky at different limiting fluxes
(i.e. {\tt \hel } -- Baldi et~al. 2002; {\tt Champ} -- Green et~al.
2004; {\tt SEXSI} -- Harrison et~al. 2003; {\tt XMM HBS} -- Caccianiga 
et al. 2004).  
The main goal of these projects is to collect a statistically significant
number ($\sim 200$) of X--ray, absorbed sources (including the ``rare'' 
QSO2) that will be used to derive the luminosity function of 
obscured objects that is nowadays basically unconstrained. 

\subsection*{Redshift and absorption distributions}
Although quite successful in reproducing the XRB spectral shape, intensity
and X--ray number counts, AGN synthesis models, at least in their
simplest version (where the evolution of the obscured population is
assumed to be the same as that of unobscured quasars), do not
reproduce other observational constraints, as the observed redshift 
and absorption distributions. \\  
There are increasing evidences (Hasinger 2003; Barger et al. 2003) 
that the bulk  of the XRB appears to be produced at relatively
low redshift (z $<$ 1) and dominated by relatively low luminosity
Seyfert galaxies (L$_{X}=10^{42}-10^{44}$ erg s$^{-1}$) rather than 
mainly due to luminous quasars at higher redshift (z=1.5-2), 
as observed by previous shallower ROSAT and ASCA surveys and 
predicted by XRB synthesis models.  \\
In the left panel of Fig.~\ref{zdist_frac} (from Hasinger 2003) 
the predictions from the Gilli et al. (2001) model are compared
with the redshift distribution of AGN selected from the CDFN and CDFS 
samples. At redshift below z=1.5, predicted and observed distributions
differ drastically, definitively indicating that the evolution of obscured 
sources is not as simple as postulated.
%
%%%%%%%%%%%%%%%%%%%%%%%%%%%%%%%
\begin{figure}[!t]
\includegraphics[width=0.47\textwidth]{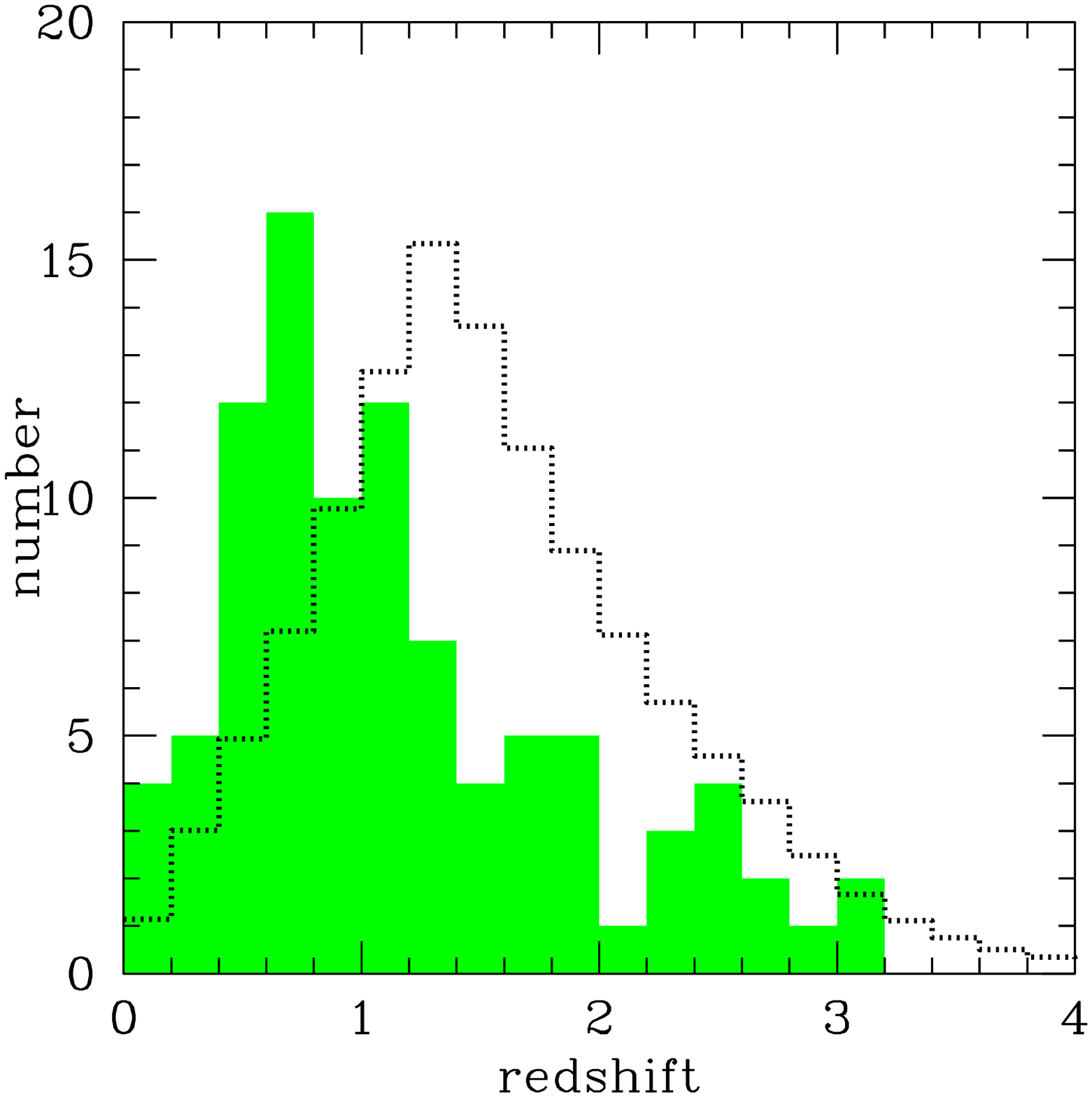}
\includegraphics[width=0.47\textwidth]{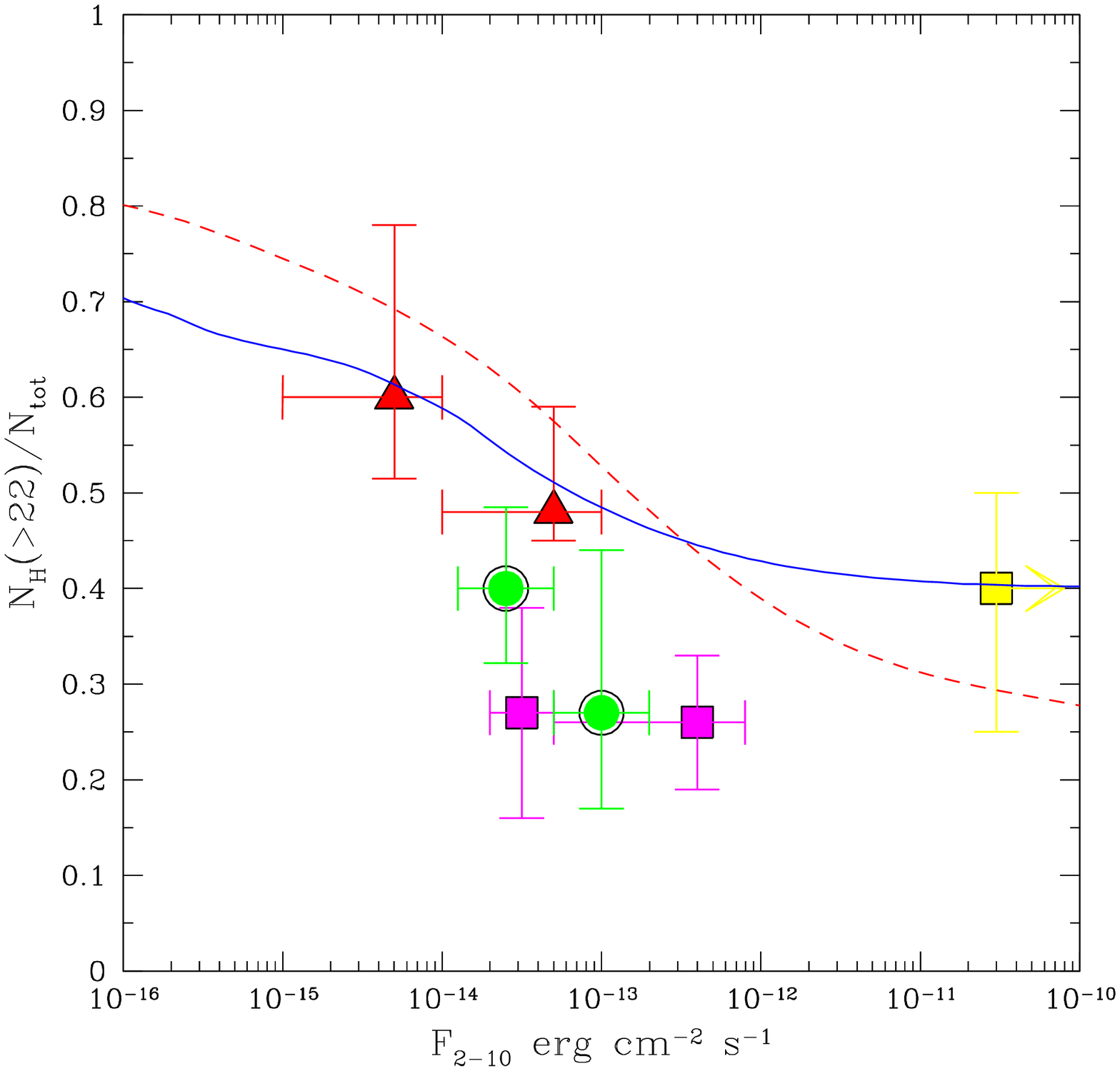}
\caption{$Left$: Redshift distribution for a selected sample of 
93 X--ray sources with $f_{2-10}>5 \times 10^{-15}$ \cgs\ 
in the central regions of the CDFS,
CDFN, Lockman Hole, Lynx field, and SSA13 field (see Gilli 2003
for details). The data (shaded area) are compared with the predictions 
of Gilli et al. (2001) at the same limiting 
flux (dotted line). [From Hasinger (2003)].
\newline $Right$: The expected fraction of objects with absorption column
densities larger than 10$^{22}$ cm$^{-2}$ as a function of the 2--10 keV
for the Comastri et al. 1995 model (solid line) and the Gilli et al.
2001 model (dashed line). The Comastri et al. 1995 model
has been normalized to the observed fraction of absorbed sources 
in the HEAO1--A2 AGN sample of Piccinotti et al. 1982 ({\it filled 
square} at bright fluxes).  
The other points and associated error bars correspond
to the the values found by Piconcelli et al. (2003, {\it filled squares})
and those obtained by Perola et al. (2004, {\it filled circles}),
both derived from proper X--ray spectral analysis.
The {\it filled triangles} at the faintest fluxes refer to the CDFN data 
(Brandt et al. 2001a; Barger et al. 2002) and have been derived from the observed
hardness ratio  assuming $\Gamma$ = 1.7, 
and the redshift proper of the source if it is available, or  $z$ = 1 otherwise.} 
\label{zdist_frac}
\end{figure}                                                                
%%%%%%%%%%%%%%%%%%%%%%%%%%%%%%%%%%%
\pn
Another important observational constraint that can be used 
to test the assumptions of the XRB models is
the number of obscured (e.g. N$_H>10^{22}$ cm\due) sources 
as a function of the X--ray flux.
While standard model predictions are able to {\it qualitatively} explain 
the hardening of the average spectrum of X--ray sources 
with decreasing flux, the rather steep hardening observed in both the CDFN 
and CDFS observations (Tozzi et al. 2001; Piconcelli et al. 2003) is a 
further indication that the evolution cannot be as simple as initially
postulated.
Moreover, preliminary results from X--ray spectral analysis 
on the fraction of obscured (N$_H>10^{22}$ cm\due) sources
indicate that observations fall short by a factor of $\sim2$ 
with respect to current model predictions at relatively bright 
X--ray fluxes ($\gtrsim 10^{-14}$, Piconcelli et al. 2003; Perola et al. 2004), 
while a much better agreement between data and predictions seems 
to occur at fainter fluxes, using the hardness--ratio technique
(Fig.~\ref{zdist_frac}, right panel). 
Finally, there are also evidences that the obscuration could be
dependent from the source luminosity and redshift (Ueda et al. 2003).\\
\pn
All the findings discussed above constitute robust evidences for 
a luminosity dependence in the number density evolution of
X--ray selected AGN, already pointed out by previous soft (ROSAT, 
Miyaji et al. 2000) and hard ({\it Beppo}SAX, La Franca et al. 2002) X--ray surveys.
Cowie et al. (2003) and Hasinger (2003) calculated preliminary 
luminosity functions (LF) on the basis of the results from the deep surveys.
The shape of the LFs in different redshift shells
is significantly different so that the cosmological evolution
can be described neither by pure luminosity nor pure density evolution.
The most surprising result is, however, that lower luminosity AGN
show much less or even negative density evolution with respect
to the strong positive evolution observed for relatively luminous QSO.
The observed evolution for the low--luminosity population 
(a very rapid increase of volume emissivity at low redshift)
can explain the sharp peak observed at z$\sim 0.7-0.8$ in the deep fields
(Hasinger 2003). \\
Interestingly enough, the late evolution of the low--luminosity AGN is
very similar to that required to explain the steep slope of  
the observed 15 micron number counts (Franceschini et al. 2001). 
Given the rather strong indications 
that the 15 micron emission is mainly due to dust enshrouded 
stellar activity (rather than AGN), a similar evolution of hard 
X--ray and MIR selected sources adds further strength 
to a close connection between the onset and fueling of AGN activity 
and star formation.
On this basis, Franceschini et al. (2002) and Gandhi \& Fabian (2003) 
elaborated XRB synthesis models able to reproduce the observed 
peak at z$\sim0.7$ in the redshift distribution, assuming 
that Type 1 AGN are distributed according to the well--determined soft 
X--ray luminosity function (Miyaji, Hasinger \& Schmidt 2000), and that 
the evolution of Type 2 AGN follows the MIR one.
However, as convincingly demonstrated by Gilli (2003), both these models 
are in disagreement with another observational constraint, the ratio
of Type2/Type1 object as a function of redshift.  \\
\pn
Although converging to the same bottom line (a luminosity-dependent cosmological evolution),
the results on the space density and evolution suffer from substantial
spectroscopic incompleteness and therefore
have still to be taken with a grain of salt.
Indeed, at the faintest fluxes explored by the CDFN and CDFS 
optical counterparts are sometimes so faint to prevent redshift 
measurements (even photometric) with the largest 
ground--based facilities; as a result, the spectroscopic 
completeness is only $\sim 50-60$\% (Barger et al. 2003; Szokoly et 
al. 2004).
This limitation mainly affects the study of the evolution of high--redshift
objects. \\

%%%%%%%%%%%%%%%%%%%%%%%%%%%%%%%%%%%%%%%%%%%%%%%%%%%%%%%
\section*{The $f_X/f_{opt}$ diagnostic}
\addcontentsline{toc}{section}{The $f_X/f_{opt}$ diagnostic}
It is well known (Maccacaro et al. 1988) that various classes of 
X--ray emitters are characterized by different values of their 
X--ray--to--optical flux ratios (hereinafter X/O)
and the observed X/O can yield important information on the nature 
of X--ray sources.  \\
For a given X--ray energy range and R--band magnitude the 
following relation holds:  
\begin{equation}
	log {\rm X/O} = log {\rm f_X} + {\rm R}/2.5 + const
\end{equation}	
\pn
where f$_{\rm X}$ is the X--ray flux, R is the optical magnitude and 
{\it const} depends only on the R--band filter used
in the optical observations\footnote{For the most popular R--band filters,
$\Delta const <=0.2$; an indicative, average value is {\it const}=5.5 
(see Hornschemeier et al. 2000) and it can be used when datasets from
different observations are compared.}.
%%%%%%%%%%%%%%%%%%%%%%%%%%%%%%%%%%%%%%%
\begin{figure*}[!t]
\includegraphics[width=\textwidth]{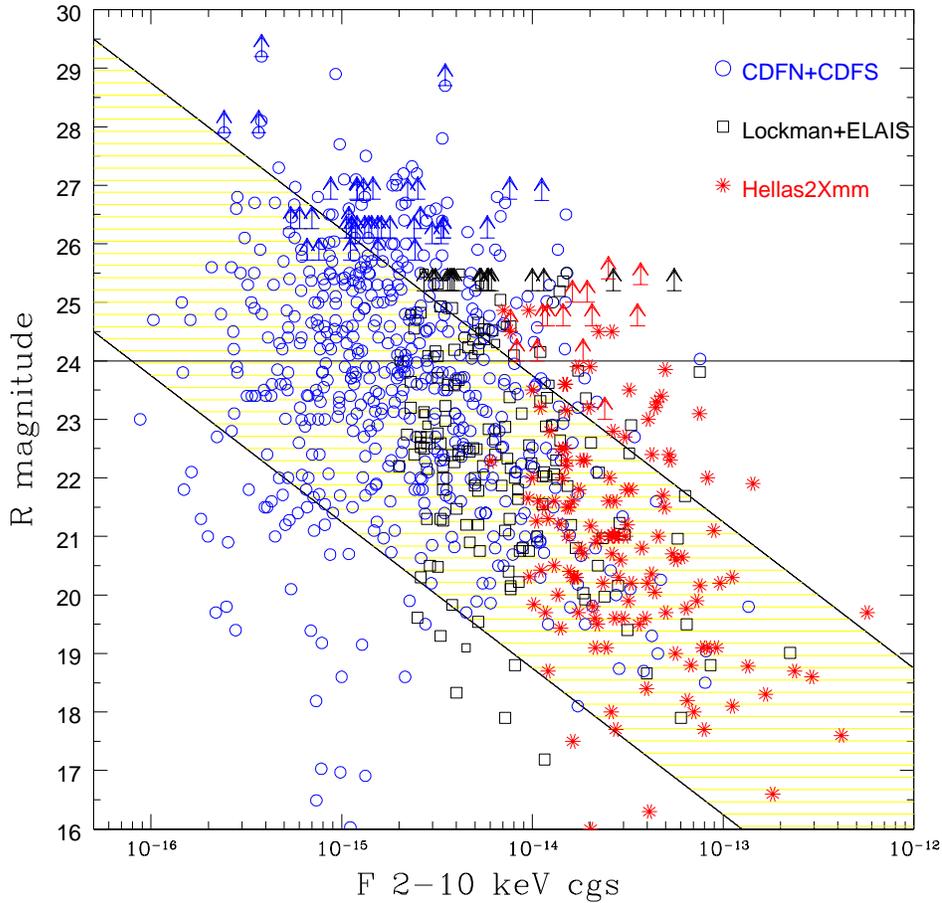}
\caption{The 2--10 keV flux versus the R band magnitude for five
different hard X--ray surveys. From bright to faint X--ray fluxes:
{\it asterisks} mark the \hel\ 
sources (this PhD Thesis); {\it squares} refer to the sources in the 
Lockman Hole \xmm\ observation (Mainieri et al. 2002) and in the 
{\rm ELAIS} deep {\it Chandra} survey (Manners et al. 2003); 
{\it circles} mark the sources detected in the {\it Chandra} Deep Field 
South (Giacconi et al. 2002) and in the {\it Chandra} Deep Field 
North (Alexander et al. 2003; Barger et al. 2003). 
The shaded area represents the region occupied by known AGN 
(e.g. quasars, Seyferts, emission line galaxies) along the correlation
log$(X/O)=0\pm 1$. The horizontal straight line at R=24 represent a 
conservative limit for spectroscopic observations.}
\label{rhx_surveys}
\end{figure*}
%%%%%%%%%%%%%%%%%%%%%%%%%%%%%%%%%%%%%%
A value of $-1 < log(X/O) < $ 1 is a clear sign of AGN activity,
since the majority of spectroscopically identified AGN 
in both ROSAT (e.g. Hasinger et al. 1998; Lehmann et al. 2001) and 
ASCA (Akiyama et al. 2000) surveys fall within this range, while normal 
galaxies and stars usually have lower ($log(X/O) <-2$) values. \\
The optical identification of sources discovered in
deep and medium deep {\it Chandra} and XMM--{\it Newton}
surveys confirms this trend to fainter X--ray fluxes and, at the same 
time, show evidence of a relatively large number of sources which deviate 
from log$(X/O) = 0\pm1$ (Fig.~\ref{rhx_surveys}).
Indeed, the range of X/O now spanned by X--ray selected sources is 
extremely large, up to 6 dex or even more. 
Most important, an interesting new population of X--ray 
sources characterized by values of their X/O$>10$ 
(i.e. sources that are optically weak with respect to the observed X--ray flux)  
is present in both deep and shallow surveys, and its fraction seems
to be constant ($\sim20$\%) over $\sim 3$ decades of fluxes. \\
Almost by definition, sources with X/O $>10$ have faint
optical magnitudes. For example, a 2--8 keV flux of 
10$^{-15}$(10$^{-16}$) \erg and an X/O=10 
correspond to $R$ magnitudes $\sim$25.5 ($\sim$28),
challenging (well beyond) the spectroscopic capabilities of 
10m-class telescopes. \\
In this respect it is important to note that shallow, hard X--ray
surveys, designed to sample relatively bright X--ray and optical fluxes, 
are best suited to investigate the nature of sources with 
high X/O: at fluxes of the order $10^{-14}-10^{-13}$ 
\erg), the magnitudes of the optical counterparts of high X/O sources 
are of the order of $R \simeq 24$ or brighter, making spectroscopic 
follow--up observations feasible.  \\
It seems reasonable to argue that most of the sources characterized 
by high X/O are high redshift, obscured AGN.
If this were the case, they could contribute to reduce the disagreement
between the redshift distribution predicted by XRB synthesis models
and that observed in deep {\it Chandra} and \xmm \ fields  
(Hasinger 2003; Gilli 2003), and provide an important
contribution to the total energy density of the background light.

%%%%%%%%%%%%%%%%%%%%%%%%%%%%%%%%%%%%
\section*{Black Hole demography} 
\addcontentsline{toc}{section}{Black Hole demography} 
An important piece of observational evidence on how  supermassive black holes (SMBHs) 
have assembled comes from the demography of the AGN population.  \\
The tight correlation observed between the black hole mass and velocity dispersion
of galactic bulges (the \msigma relation, Gebhardt et al. 2000;
Ferrarese \& Merritt 2000) is generally taken as strong
evidence that the growth of SMBH and the formation of galaxies go 
hand in hand (see, e.g., Ho 2004).
The first implication of this relation is that most (if not all) 
galactic bulges contain  ``dormant'' SMBHs. 
If this is the case, the observed local black hole mass density %in remnant black holes 
($\rho_{\bullet}$) should be explained in terms of the overall 
black hole mass density accreted in the AGN phase.
The BH mass density due to AGN can be estimated from 
the observed radiation($S$) %energy density 
emitted by active black holes in a given energy
band over the cosmic time, using the elegant Soltan's argument (Soltan 1982): 
%%%%%%%%%%%%%%%%%%
\begin{equation}
\rho_{\bullet} = {1 \over \eta c^2} {4\pi k_{bol} \over c} (1+<z>) 
\int S\ \frac{dn}{ds}\ dS. 
\label{soltan}
\end{equation}
%%%%%%%%%%%%%%%%% 
that, besides the knowledge of %the radiation ($S$) emitted in a given energy band, 
the differential counts ($\frac{dn}{dS}$) and of the average redshift 
of the peak of AGN activity ($\langle z \rangle$), requires 
{\it assumptions} about the efficiency ($\eta$) of turning accreted 
rest-mass energy into radiated energy, and the bolometric correction 
($k_{bol}$) which relates the total emissivity to the emissivity 
in the chosen band. \\
\pn
Hard X--ray surveys have already probed the largest fraction of the whole 
AGN population; this makes the results obtained from both deep
and shallow surveys the most suitable to estimate the {\it total} black 
hole mass density due to accretion.
Moreover, given that the XRB spectral intensity records the bulk of the 
accretion power, the integral in Equation~\ref{soltan} can be 
rewritten:
%%%%%%%%%%%%%%%%%%% 
\begin{equation}
\rho_{\bullet} =
{1 \over \eta c^2} {4 \pi k_{bol} \over c} (1 + \langle z \rangle) I_0  
\label{bhformula}
\end{equation}
%%%%%%%%%%%%%%%%% 
where $\langle z \rangle$ is the average redshift of the X--ray sources detected 
in the 2--10 keV band and $I_0$ is the absorption corrected XRB intensity in the
same energy band. \\
Early estimates, made assuming the same evolution 
for obscured and unobscured sources (e.g. $\langle z \rangle$=2 observed
for soft X--ray selected QSO) and coupled with the evidences that more than 
80\% of the accretion is obscured, %(Fabian \& Iwasawa 1999), 
led to a $\rho_{\bullet}=6\div 17\times 10^{5}$ M$_{\odot}$ Mpc$^{-3}$, 
if a standard $\eta$=0.1 is adopted (Fabian \& Iwasawa 1999; 
Elvis, Risaliti \& Zamorani 2002). \\
With the data on hard X--ray selected active SMBHs rapidly accumulating, 
a lower value for $\rho_{\bullet}$ from AGN 
appears more plausible (see Fabian 2003 for a recent review). 
Moreover, now it is possible to safely (e.g. with the lowest number of arbitrary assumptions)
compare the black hole mass density expected from the AGN activity with that 
observed in nearby galactic bulges (e.g. Haehnelt 2003).

%%%%%%%%%%%%%%%%%%%%%%%%%%%%%%%%%%%%%%%%%%%%%%%%%
\section*{Do Unified Schemes work?}
\addcontentsline{toc}{section}{Do Unified Schemes work?}
As already shown, current synthesis models, built on a strictly X--ray based scheme, 
are far from being unique (see Comastri 2001 for a
review); in particular, a simultaneous fit to all the observational
constraints requires {\it at least} to relax some of the key assumptions
in Unified Schemes.
Moreover, the broad band Spectral Energy Distribution (SED)
of the XRB constituents, outside the X--ray domain, is essentially
unconstrained, with the consequent lack of model predictive power at
longer wavelengths. 
In this respect it is not surprising that the
results of multi--wavelength follow--up from both shallow and deep surveys, 
have started to unveil that the sources responsible for a large fraction 
of the XRB energy density are characterized by broad band properties 
which are significantly different from those of  AGN selected in the optical 
and soft X--ray bands (Barger et al. 2002, Giacconi et al. 2002; Willott et al. 2003). 
Although there are compelling theoretical and observational evidences  
which suggest that the large majority of the 
hard X--ray sources are obscured AGN, the origin of such a broad variety 
in their multiwavelength properties is still far to be understood.  \\
On the one hand, there is rather increasing evidence of the
presence of luminous X--ray sources in the nuclei of 
galaxies without any
evidence of optical nuclear activity nor of a high star formation
rate in their optical spectra, which are 
instead typical of early-type ``normal'' galaxies (Hornschemeier et al. 2001;
Giacconi et al. 2001; Barger et al. 2001; Comastri et al. 2002a,b). 
Moreover, there are evidences that the one--to--one
relation postulated in Unified schemes between optical Type 1 and X--ray 
unobscured sources, and between optical Type 2 and X--ray 
obscured sources, does not 
hold expecially at high redshifts and luminosities (Brusa et al. 2003; Page et al. 2003). \\
On the other hand, the hard X--ray selection turned out to be very efficient 
in revealing an AGN population with optical to near--infrared colours 
redder than those of optically selected QSOs.
In this respect, the discovery that a sizable fraction of hard X--ray sources
are also associated to extremely red objects (EROs) with optical to near--infrared
colour R-K $>$ 5
(Lehmann et al. 2001; Mainieri et al. 2002; Alexander et al. 2002)
is even more intriguing.  
The observed optical to near-infrared colors
of EROs are consistent with both passively evolving elliptical
galaxies at z$\sim$1 or with dust reddened starburst galaxies and obscured 
AGN. Given the key role played by this class of objects to constrain
models for the formation and evolution of massive elliptical galaxies
and star formation at high redshift, X--ray observations of those 
objects provide an exciting opportunity to investigate the link between
galaxies and AGN evolution. 

\vspace*{1.6truecm}
\begin{flushleft}
\chapter*{The goals of this PhD Thesis}
\end{flushleft}
\addcontentsline{toc}{chapter}{\bf The goals of this PhD Thesis}
\vspace {1.0truecm}

\pn
The main goal of the present work is to provide further contributions 
to the understanding of the physical and  evolutionary properties of AGN,
which are %i.e. the basis of {\it observational constraints} 
necessary to derive the accretion history of super massive black holes that 
reside in most of galaxies. \\
\pn
In order to make significant progress in this direction the
following issues will be addressed:
\begin{itemize}
\item[A)] What is the nature and luminosity/redshift dependence of 
 sources which seem to deviate from the simplest version of unified models? 
 How many are they? 
\item[B)] What is the nature of X--ray bright optically faint sources? 
 What is their redshift? Are they highly obscured, high redshift AGN? 
\item[C)] Which is the the evolution of hard X--ray sources? Is the evolution
 of low--luminosity AGN different from that of high--luminosity AGN?
 Is the relic Black Hole mass density accreted in the active phase 
 consistent with the local one?  
\item[D)] What are the optical/IR colors of obscured hard X--ray sources? 
 How many hard X--ray sources are extremely red objects (EROs)? 
 Conversely, what are the high energies properties of the near--infrared 
 selected ERO population? How many infrared selected EROs are X--ray emitters? 
\end{itemize}
\pn
Observations at high energies  yield important information
on the structure and nature of AGN; when coupled with deep
optical and near--infrared (photometric and spectroscopic) follow-up, 
they provide constraints on the mass of the
growing black holes and, therefore, are essential 
to better understand the nature of the various components of the
X--ray background light and can be used as test for the
accretion paradigm.% (L$_{\rm Edd}$, energy release etc.). 
In this framework, we have started a program of multiwavelength
follow-up observations of hard X--ray selected sources
serendipitously discovered in XMM--{\it Newton} fields
over $\sim 4$ deg$^{2}$ (the \hel\ survey; Baldi et al. 2002).\\
Conversely, optical and near--infrared surveys of galaxies are crucial
to discriminate between different cosmological scenarios
(e.g. hierarchical or monolithic growth of the structures) and, thus,
to recover the galaxy evolution path.
With a complementary approach to that of hard X--ray surveys, in order to investigate 
the link between nuclear activity and the galaxy formation,
\xmm\ and \chandra\ observations of photometric and spectroscopically 
selected EROs have been obtained.  \\
\pn
The \hel\ survey will be described in Chapter 1, with particular
emphasis on the multiwavelength data obtained for the sources
detected in $\sim 1$ deg$^{2}$, and on 
the importance of high spatial resolution observation in the optical 
identification process of hard X--ray sources counterparts. \\
In Chapter~2 I will present the multiwavelength
properties of hard X--ray sources, as revealed from optical and X--ray 
spectroscopy of the brightest population responsible 
for the XRB energy density. In particular, I will focus the discussion
on those sources which seem to deviate from the simplest version of unified models. \\
In Chapter~3 I will present spectroscopic identifications and near--infrared
analysis of objects with high X/O selected in the \hel\ survey. I will
also present different approaches developed to estimate the redshift 
of optically faint X--ray sources and needed to build an almost ``complete''
sample of hard X--ray selected sources. \\ %the redshift distribution 
In Chapter~4 the results from the \hel\ survey, complemented
with those obtained from the deep surveys, will be used to derive
the number and luminosity densities as a function of redshift.
Observational constraints for the cosmic evolution, the integrated
black hole mass density and the QSO2 will also be discussed. \\
In Chapter~5 the X--ray and optical properties of X--ray detected 
EROs from a large and complete near--infrared sample 
are presented, and compared with those of hard X--ray sources with
similar R-K colors and QSO2. A selection criterion to pick up QSO2
on the basis of the observed optical, near--infrared and X--ray fluxes
is proposed. \\
As a comparison with the results on X--ray detected EROs, 
in Chapter~6 the average high--energy properties of 
non--AGN EROs are investigated through the ``stacking analysis''
technique, and the existence of a dichotomy in the spectroscopic
classification is confirmed also in the X--rays. \\
Finally, in the last Chapter, I will summarize the most important
results of the present work and I will briefly discuss the 
implications for future observations. \\
I will assume a $\Lambda$CDM cosmology, with the following values
for the Hubble constant and the cosmological parameters:
H$_0$=70 km s$^{-1}$ Mpc$^{-1}$, 
$\Omega_{\Lambda}$=0.7, $\Omega_{\rm m}$=0.3 (Spergel et al. 2003).

\clearpage
\markboth{\sc \ }{\sc Conclusions}

\addcontentsline{toc}{chapter}{\bf Conclusions}
\vspace*{1.6truecm}
\begin{flushleft}
{\Huge \bf Conclusions}
\end{flushleft}
\vspace {1.0truecm}
\setcounter{chapter}{8}
\setcounter{section}{0}
\setcounter{figure}{0}
\setcounter{table}{0}
\setcounter{equation}{0}

\pn
The most important results obtained in this PhD Thesis concern
the physical and evolutionary properties of obscured AGN 
detected in hard X--ray surveys, that contributes most to the 
accretion history in the Universe. \\
\begin{itemize}
%%%%%%%%%%%%%%%%%%%%%%%%%%%%%%%%%%%%%%%%%%%%%%%
%
%  HIGH RESOLUTION
%
\item The \hel\ survey 
has provided optical identifications and 
spectroscopic classifications for 97 out of 122 hard X--ray selected 
sources detected in $\sim$1 deg$^2$ (the \odf\ sample).
The spectroscopic completeness of the sample ($\sim$80\%, mainly 
limited by the faintness of the optical counterparts) is one
of the highest for sources detected at flux level $\gtrsim 10^{-14}$
\cgs\ and the results from the \hel\ survey 
represent a complementary and necessary probe of the
hard X--ray sky with respect to the deep, pencil--beam \chandra\ and \xmm\ 
surveys. \\
A detailed study of one specific field (the PKS~0312--77 field) 
for which \chandra, radio and near--infrared data were 
also available, has clearly demonstrated the need for  
high spatial resolution observations to exactly identify the hard 
X--ray sources counterparts (Chapter 1). 
%
% MULTIWAVELENGTH
%
\item The overall picture emerging from the optical identifications 
of the \odf\ sample indicate 
a wide spread in the optical (both in the continuum shape and 
emission lines) and X--ray properties of the sources
responsible for the bulk of XRB energy density, confirming and extending 
results obtained from both deep and shallow surveys. 
\begin{itemize}
    \item The combination of X--ray spectral analysis and deep VLT spectroscopy has revealed that $\sim 10$\% of high--redshift, high--luminosity objects optically classified  as unobscured BL AGN are absorbed in the X--rays  
by column densities in excess than $10^{22}$ cm$^{-2}$. \\
    \item The elusive properties of those AGN classified as X--ray Bright Optically
Normal Galaxies ({\tt XBONG}) 
can be due to a combination of the absorption associated with the AGN and 
the optical faintness of the nuclear emission with respect to the host galaxy.
Obscuration of the nuclear source by large columns (possibly Compton thick)
of cold gas seems to provide the most plausible explanation of the observed 
broad--band properties, though such a possibility is not unique. 

\end{itemize}
The optical appearance of hard X--ray selected AGN is different from 
what expected on the basis of classic Unified Schemes, calling for some %that therefore need some 
revisions to account for the discrepant classifications 
produced by the properties of the circumnuclear medium, 
and/or by geometrical or beaming effects (Chapter 2).
%
% HIGH X/O and OTHER REDSHIFTS measurements
%
\item Thanks to the large area covered, 
we have obtained the first spectroscopic identification 
of a sizable sample of objects 
with an extreme X--ray to optical flux ratio (X/O $>10$). \\
Different approaches, beside the photometric technique,
to estimate the redshift of optically faint X--ray sources, 
have also been developed and tested to derive the 
redshift distribution of highly obscured, even Compton thick AGN: 
\begin{itemize}
\item {\sc colour-based redshifts}: 
coupling the morphological information derived from K--band surface brightness profiles
for a subsample of X/O$>10$ sources in the \hel\ survey, with the observed R-K colours, 
it was possible to derive a ``minimum'' redshift for these objects; 
\item {\sc a statistical approach}: on the basis of the optical to X--ray properties
 of identified sources it was possible to statistically assign luminosities --- hence the
 redshifts --- to optically faint X--ray sources in a combined sample of hard X--ray selected objects.
\item {\sc the detection of strong FeK$\alpha$ features}: 
it was possible to derive the redshift of highly obscured objects by detecting a strong FeK$\alpha$ line
in a few high S/N X--ray spectra of sources detected in the deepest \chandra\ fields.
\end{itemize}
A detailed comparison with X-ray selected sources in various deep and medium--deep surveys 
indicates that heavy (N$_H > 10^{22}$ cm$^{-2}$) obscuration is 
almost ubiquitous among objects with high X/O and that obscured sources 
(in particular QSO2, the high-luminosity, high-redshift 
obscured AGNs predicted in XRB synthesis models) 
can be hosted in the bulge of luminous, massive ellipticals which already formed the bulk 
of their stars at high redshift (Chapter 3).
%
% EVOLUTION + BH MASS
%
\item Using the correlation observed between the X/O and the
X--ray luminosity for  obscured sources, it was possible 
to build a ``virtually complete'' sample of identified hard X--ray 
sources over a wide range of redshifts and luminosities. 
We have confirmed that a luminosity-dependent density evolution 
for the sources responsible for the XRB (low luminosity sources 
peaking at a later cosmic time) is needed to match the observed
number and luminosity densities as a function of redshift. 
With this assumption, and the hypothesis that hard X--ray surveys
probe the largest fraction of the whole AGN population,
it is possible to explain the observed local BH mass density
as entirely  due to the growth of AGN (Chapter 4). 
%
% EROS AND QSO2
% 
\item 
The first comprehensive characterization of the
X--ray properties of a large sample of X--ray detected
EROs has been presented. Results obtained from a 80 
ks XMM--{\it Newton} observation 
indicate that, at the relatively bright X--ray and
near--infrared fluxes probed by the present observation, 
AGN contribute only for a negligible fraction ($\sim$3\%) to the
optically selected EROs population. 
Although a spectroscopic redshift is not available for all of the
sources, the X-ray, optical, and near-infrared properties of X-ray
selected EROs nicely fit those expected for QSO2, 
confirming the results obtained in Chapter 3 from the \hel\ survey
(Chapter 5). 
%
% EROS K20
%
\item The results of \chandra\ stacking analysis of a well defined sample of
spectroscopically identified, non--AGN EROs suggest that the
dichotomy in the spectroscopic classification of the majority of K--selected, non--AGN EROs 
(e.g. dusty and old systems) appears to be present also in the X--rays:
``dusty'' objects are relatively bright X--ray sources
(L$_X\sim10^{41}$ erg s$^{-1}$), while ``old'' EROs are below the
detection threshold. 
Moreover, using the hard X--ray luminosity as a Star Formation Rate
(SFR) indicator,  it has been possible to estimate an average SFR
for the dusty population in the range 5-44 M$_{\odot}$ yr$^{-1}$;
this estimate is lower than, although consistent with, the value based
on the reddening--dependent [OII] emission and is in agreement with
results from recent far--infrared and radio observations (Chapter 6).
\end{itemize}
%%%%%%%%%%%%%%%%%%%%%%%%%%%%%%%%%%%%%%%%%%%%
Undoubtedly, the somewhat unexpected link between EROs and QSO2, 
is intriguing: near infrared observations of obscured QSO 
selected on the basis of their high X/O and, conversely, hard 
X--ray observations of a complete sample of EROs, constitute 
one of the strongest evidences that these two populations originally 
discovered at different wavelengths are intimately connected. 
{\bf A selection criterion based on the X/O and the R-K 
colour of hard X--ray selected sources has been proposed, to efficiently 
pick--up the elusive QSO2 population}, difficult to select at optical 
wavelengths.
Furthermore, X--ray detected EROs can be used as lighthouses to 
investigate the accretion paradigm at high redshifts to address the 
issue of elliptical galaxy formation and the expected co-evolution with the accreting black-holes. 

\section*{Future Directions}
The description of the nature of hard X--ray selected 
sources has already approached a fairly complicate picture.
While the general trends are rather robust (several
independent arguments converge to the similar qualitative results) the
number of X-ray sources with spectroscopic identifications is still far too
small to derive accurate quantitative information about the
AGN evolution at high redshift (z=2-4). 
Moreover, in order to obtain robust estimates of the
black hole mass density due to growth by accretion,
a much better knowledge of the redshift, luminosity
and Spectral Energy Distributions (SEDs, related to $k_{bol}$)
of highly obscured AGN is of paramount importance. \\
The absorbed radiation emitted by the nucleus is expected to be re-emitted 
in the far--infrared and sub--mm bands. 
In this framework, {\it Spitzer} observations (from 
3.6 $\mu$m to 170 $\mu$m) of a well--selected sample of sources 
detected in the \odf\ sample (including {\tt XBONG}, high X/O, 
EROs and QSO2) have been proposed, to complement the 
multiwavelength database and to probe the physical properties of 
those AGN for which the presence of SMBH is inferred from 
the X--ray emission but which remains elusive from 
observations at optical--UV and near--infrared wavelengths.
Moreover, SCUBA (850 $\mu$m) observations of an X--ray selected 
sample of X--ray Type II QSO could be crucial to derive a 
solid estimate of the bolometric luminosity of accretion 
powered hard X-ray sources and of the correlations between 
X--ray luminosity and obscuration with infrared emission.  \\
\pn
The results from both deep and shallow surveys have unambiguously
unveiled a differential evolution for the low-- and high--luminosity 
AGN population. 
In this respect, the evolution of the obscured AGN luminosity function
remains the key parameter to be determined. 
A study takling simultaneously the problems of the 
shape and evolution of the luminosity function, and of the N$_H$ 
distribution as a function of luminosity and cosmic epoch 
(with an approach akin to that followed by Ueda et al. 2003), 
will be the subject of future investigations (La Franca et al., 
in preparation).\\
The best strategy to properly address this issue is to increase 
the area covered in the X--ray band, and the corresponding 
optical-NIR photometric and spectroscopic follow-up, down to 
F$_{2-10keV}=1-5\times10^{-15}$ \cgs, where the bulk of the 
XRB is made, rather than pushing the depth of the survey
beyond the present limits of the deepest \chandra\ surveys. \\
%However, the sample sizes are still insufficient to attempt deriving the 
%evolution of obscured AGN {\it separately} from that of unobscured AGN.
On the one hand, the optical follow-up of the \hel\ sources will be 
extended from 1 to 4 deg$^{2}$, in order to obtain a sample of 100-150 
obscured objects at F$_{2-10keV}>10^{-14}$ \cgs, sufficient 
to constrain  the high--luminosity tail of their X--ray luminosity 
function and to adequately figure their differential 
evolution over 2-3 luminosity dex and up to z$\sim$2 
(Cocchia, PhD Thesis). \\ 
On the other hand, the \xmm\ survey of the ELAIS--S1 field
(Puccetti, PhD Thesis) will push the limiting fluxes down to 
$2\times 10^{-15}$ \cgs\ over $\sim0.5$ deg$^{2}$ 
allowing to considerably increase the statistic on high redshift, 
moderate luminosity objects.\\ 
Finally, a comprehensive multiwavelength program is underway to observe 
a contiguous area of $\sim 2$ deg$^{2}$ (the ``COSMOS'' project) using 
all the modern (HST/ACS, {\it Spitzer}, \xmm, \chandra, VLT/VIMOS, VLA, GALEX)
and future (e.g. ALMA) observatories from radio to X--rays. 
These data will provide the necessary
photometry to characterize the SEDs of some thousands of 
AGNs as a function of redshift and type, and to explore almost all the aspects of
galaxy/AGN evolution. 
The large area covered will allow a detailed comparison
between galaxy and AGN clustering, and theoretical models for the
evolution of the large scale structure of the Universe. This will shed
new light on the issue of whether AGNs trace higher density peaks or
higher mass haloes in the high--redshift universe and how active sources 
are correlated with the environment.  \\
\pn
With the full exploitation of all the multiwavelength data, it will be
possible to build up large and homogeneously--selected samples of
almost all the classes of objects revealed in the deep imaging
surveys (e.g: AGN, EROs, B--dropout, High--z star-forming galaxies, 
cluster of galaxies etc.).\\
In particular, the combination of optical, near--infrared and 
X--ray (both \chandra\ and \xmm) data of complete samples of EROs 
are crucial to definitively assess the fraction of reddened
sources among the XRB constituents, the link between EROs,
high X/O and QSO2, and the r\^ole of these objects in the
framework of galaxy formation.
%
% FUTURE
%

\end{document}